\documentclass[reprint,aip,pop,amsmath,amssymb,showpacs,nofootinbib]{revtex4-1}

\usepackage{graphicx}	% for eps images
\usepackage{eucal}		% better fonts for \mathcal
\usepackage{mycommands}

\newcommand{\vk}{\v{k}}

\newcommand{\kbar}{\ol{k}}
\newcommand{\kbsq}{\kbar^2}

\begin{document}

\title{Numerical simulation of the geometrical-optics reduction of CE2 and comparisons to quasilinear dynamics}

\author{Jeffrey B.~Parker}
\email{parker68@llnl.gov}
\affiliation{Lawrence Livermore National Laboratory, Livermore, CA 94550}

\begin{abstract}
Zonal flows have been observed to appear spontaneously from turbulence in a number of physical settings.  A complete theory for their behavior is still lacking.  Recently, a number of studies have investigated the dynamics of zonal flows using quasilinear theories and the statistical framework of a second-order cumulant expansion (CE2).  A geometrical-optics (GO) reduction of CE2, derived under an assumption of separation of scales between the fluctuations and the zonal flow, is studied here numerically.  The reduced model, CE2-GO, has a similar phase-space mathematical structure to the traditional wave-kinetic equation, but that wave-kinetic equation has been shown to fail to preserve enstrophy conservation and to exhibit an ultraviolet catastrophe.  CE2-GO, in contrast, preserves nonlinear conservation of both energy and enstrophy.  We show here how to retain these conservation properties in a pseudospectral simulation of CE2-GO.  We then present nonlinear simulations of CE2-GO and compare with direct simulations of quasilinear (QL) dynamics.  We find that CE2-GO retains some similarities to QL.  The partitioning of energy that resides in the zonal flow is in good quantitative agreement between CE2-GO and QL.  On the other hand, the length scale of the zonal flow does not follow the same qualitative trend in the two models.  Overall, these simulations indicate that CE2-GO provides a simpler and more tractable statistical paradigm than CE2, but CE2-GO is missing important physics.
\end{abstract}

\maketitle

\section{Introduction}
Zonal flows in fluids are of fundamental physical interest.  These flows, which alternate in space and are often quasistationary in time, form spontaneously in the disorder of turbulent flow and persist as a coherent structure.  Such flows have been observed in plasmas, planetary atmospheres, and even in simulations of astrophysical discs.\cite{fujisawa:2009,hillesheim:2016,vasavada:2005,johansen:2009,kunz:2013}  In magnetically confined fusion plasmas, zonal flows have been linked to the regulation of turbulence and are thought to play a critical role in determining the overall level of heat transport.\cite{lin:1998,diamond:2005}  As heat loss is one the key parameters determining overall performance of a fusion reactor, zonal flows have attracted much attention.

A strong theoretical understanding of zonal flows in fusion plasmas is therefore of interest.  Realistic descriptions of magnetically confined plasmas are complicated, and simulations will ultimately be needed for quantitative investigations of zonal flows.  Already, turbulence simulations have uncovered important modifications to zonal flow arising from multiscale interactions between ion-scale and electron-scale fluctuations.\cite{howard:2016a,howard:2016b}

On the other hand, the fundamental theory of zonal flow is a relatively unexplored area.  Analytical approaches can have a crucial role in providing both physical insight and conceptual frameworks with which to interpret data or simulations.  The problem of studying zonal flow \emph{and} turbulence seems overwhelming at first glance, when turbulence itself is practically intractable on its own, but there are several reasons to think progress could be made.  First, the basic issues under consideration---why do zonal flows form, what sets their overall length scale and amplitude, how do they interact with turbulence---are much less intricate than questions about small-scale inertial ranges, dissipation, or intermittency.  Second, \emph{inhomogeneous} flow with coherent structures whose mathematical description is relatively straightfoward, such as zonal flows, may actually be \emph{more} tractable because dynamics at large scales have greater structure than in homogeneous, isotropic turbulence.  Finally, the ubiquity of zonal flows suggests that the nature of the physics determining zonal flow does not depend on the details of the turbulence per se, and that crude models representing the turbulent fluctuations could provide a foothold for gaining insight into the basic physics.

Some theoretical approaches to understanding zonal flows include potential vorticity staircases\cite{dritschel:2008,scott:2012} and wave-kinetic theory\cite{manin:1994,smolyakov:2000b}.  Wave-kinetic theory adopts a statistical viewpoint, considering a second-order statistical average over the fluctuations and its interactions with zonal flow.  The wave-kinetic approach has been advocated as being intuitively appealing because of its Vlasov-like kinetic description involving ray trajectories of wavepackets through phase space as well as the capacity to understand dynamics through conservation of action.  However, it was recently pointed out that the wave-kinetic equation traditionally used in the literature has some serious defects.\cite{parker:2016,ruiz:2016}  First, it does not conserve enstrophy.  Second, the dynamics are dominated by growth of arbitrarily small scales of zonal flow, as confirmed by numerical simulations in which the zonal flow occupies the highest resolved wavenumbers.

We consider a similar statistical approach that addresses the weaknesses of the wave-kinetic theory.  This recent line of work begins from the so-called second-order cumulant expansion (CE2); an equivalent approach to CE2 is known in the literature as SSST or S3T.\cite{farrell:2003,farrell:2007,bakas:2013b,bakas:2015,constantinou:2014,constantinou:2016,srinivasan:2012,marston:2008,tobias:2011,parker:2013,parker:2014,squire:2015}  One motivation for CE2 is to strip away inessential details in the hopes of understanding the crucial elements for zonal flow as well as making analytic progress more tractable.  One way to derive CE2 can be understood through the quasilinear (QL) approximation.  In a decomposition of fields into a mean component (zonal flow) and fluctuation component (also referred to as the drift wave or eddy component), the QL approximation neglects fluctuation-fluctuation nonlinearities affecting the fluctuations and retains the fluctuation-mean nonlinearities.  In other words, in Fourier space, all triads consisting of three drift-wave modes are neglected, whereas triads consisting of two drift-wave modes and a zonal-flow mode are retained.  While this approximation may be justified when the zonal flow is strong, we adopt here the point of view that the QL approximation is also valuable for understanding the qualitative behavior rather than the quantitative details.  The QL approximation is certainly a drastic truncation, and is not expected to be quantitatively correct in all situations.  On the other hand, some authors have explored asymptotic limits where QL may be justified.\cite{bouchet:2013}  Truncations intermediate between the original equation and the QL approximation are also under investigation.\cite{marston:2016}

Following the QL approximation, a straightforward statistical averaging yields the CE2 equations.  The standard \emph{closure problem} of statistical equations, where one requires unknown triplet correlations, does not appear in the QL system.  Hence, one can obtain a closed system.  The only additional assumption required is an ergodic assumption that a zonal average is equivalent to an ensemble average.  This assumption becomes better satisfied as the domain size in the zonal direction increases.

CE2 consists of a set of coupled equations for the fluctuations and the zonal flow which retain quadratic nonlinearity.  There are evolution equations for the two-point \emph{covariance} of the fluctuations equations and the one-point \emph{amplitude} of the zonal flow.  Statistical realizability of CE2 is guaranteed because it is the exact statistical description of the quasilinear system.  This is a nontrivial feature because statistical closures often face difficulties with statistical realizability, leading to unphysical behavior such as negative energies.\cite{krommes:2002}

A substantial amount of analytic progress has been made in understanding the behavior of zonal flows within CE2.  Such analytic work can contribute to conceptual frameworks for understanding zonal flow behavior, and may be useful even outside of the QL approximation.  In recent years, a theoretical framework for zonal flow formation and equilibration has emerged.\cite{srinivasan:2012,parker:2013,parker:2014}  First, formation of zonal flow has been understood as a symmetry-breaking instability of a statistically homogeneous state known as zonostrophic instability.  That is, the statistical equations allow for a steady-state solution that is spatially homogeneous and does not have zonal flow.  But this steady state can be zonostrophically unstable, giving rise to new stable solutions with a spontaneously broken symmetry---statistical homogeneity---and exhibit zonal flow.  CE2, unlike the wave-kinetic theory, makes no assumption of scale separation between zonal flow and fluctuations, and the zonostrophic instability is valid for all wavelengths of the zonal flow.  The zonostrophic instability is also related to the so-called secondary instability.  Secondary instability describes the tendency for a ``primary'' eigenmode to be unstable to a perturbation and its sidebands.\cite{rogers:2000,plunk:2007,pueschel:2013,lorenz:1972,gill:1974}  It has been shown that the dispersion relation of a primary eigenmode to a secondary zonal flow is identical to the dispersion relation of zonostrophic instability in the appropriate limit.\cite{parker:thesis}  Thus, the zonostrophic instability can be thought of as a generalized modulational instability whereby an entire fluctuation \emph{spectrum} is unstable to a regular coherent structure.

Beyond the linear growth stage, one is interested in the eventual equilibration of zonal flow.  A detailed perturbation expansion within CE2 about the marginality point of zonostrophic instability shows that just above criticality in the weakly nonlinear regime, zonal flows obey a real Ginzburg--Landau equation.\cite{parker:2013,parker:2014}  That realization connects the physics of zonal flows to the wider field of pattern formation, which provides many broad insights into the behavior of physics involving spontaneously broken symmetries.\cite{cross:1993,cross:2009}  For example, understanding the length scale of zonal flow in the nonlinear regime can be understood through stability boundaries that bracket a range of stable wavelengths.

In this work, we explore the CE2-geometrical-optics (CE2-GO) model, which is a geometrical-optics (GO) reduction of CE2.  Previous work has described the relation of CE2-GO to CE2 and to the traditional wave-kinetic equation.\cite{parker:2016,ruiz:2016}  It was shown that in at least some regimes, the linear growth rates of zonostrophic instability for CE2-GO are nearly identical to those in full CE2 (or equivalently, QL) dynamics.  The contribution of this paper is to investigate the CE2-GO model in more detail through nonlinear simulations, using a conservative, pseudospectral scheme.  In particular, we compare these simulations with QL simulations in order to assess the fidelity of the GO approximation.  We find that some aspects of QL dynamics are properly retained, such as the partitioning of the fraction of total energy into the zonal flow.  On the other hand, some physics appear to be lost in the GO approximation, namely the processes that set the final length scale of the zonal flow.

We reiterate that our perspective is not that the nonlinear eddy-eddy terms neglected in the quasilinear approximation are unimportant.  Rather, we take the point of view that physical insight can be gained by studying zonal flow in the simplest models possible.  In addition, a more complete statistical closure beyond the QL approximation would keep some representation of the physics of the eddy-eddy interactions.  In such a statistical description, the eddy-zonal flow interaction, which is represented exactly within CE2, could be simplified with the same GO approximation as is used in CE2-GO.  The first step towards that goal is to verify the usefulness of the GO reduction.  In that endeavor, the appropriate comparison for CE2-GO is with the quasilinear dynamics rather than the full nonlinear dynamics; that is what we undertake here.  In Section \ref{sec:CE2-GO}, we present the CE2-GO model.  In Section \ref{sec:numerical_simulation}, we formulate a conservative pseudospectral scheme for numerical simulations.  In Section \ref{sec:results}, we present results of simulations of both CE2-GO and the QL model.  In Section \ref{sec:discussion}, we discuss how the CE2-GO and the QL simulation results compare and what the consequences are for the fidelity of the CE2-GO model, and in Section \ref{sec:conclusion} we offer our conclusion.

\section{CE2-GO model}
\label{sec:CE2-GO}
As our paradigm model, we use the 2D Modified Hasegawa--Mima equation,\cite{smolyakov:2000, krommes:2000} 
	\begin{gather}
		\partial_t \z + \v{v} \cdot \nabla \z + \b \partial_x \psi = f + D, \label{HMEquation} \\
		\z = \bigl( \nabla^2 - \hat{\a} \bigr) \psi, \label{HMPoissonEqn}
	\end{gather}
where $\z$ is the generalized vorticity, $\psi$ is the electric potential, $\v{v} = \unit{z} \times \nabla \psi$ is the $\v{E} \times \v{B}$ velocity, $\b$ is the inverse density scale length, and $f$ and $D$ represent forcing and dissipation.  Lengths are normalized to the sound radius $\r_s$ and times are normalized to the drift-wave period $\w_*^{-1} = (L_n / \r_s) \W_i^{-1}$.  In Eq.~\eref{HMPoissonEqn}, $\hat{\alpha}$ is an operator that is zero when acting on zonally averaged modes $(k_x=0)$ and one otherwise $(k_x \neq 0)$, which ensures appropriate adiabatic-electron dynamics.  As the Hasegawa--Mima equation has no intrinsic instability that provides excitations of fluctuations, $f$ is added as an external white-noise forcing, similar to a stirring.  Dissipation $D$ is then required to balance the external energy input and allow for a statistical steady state.  For simplicity and tractability, we let the dissipation consist of a linear drag $\m$ and hyperviscosity $\n$.  The geophysical coordinate convention has been used, where the density gradient and the velocity of the zonal flow vary in the $y$ direction, and the zonal direction is along $\unit{x}$.

From the equation of motion, the statistical CE2 equations can be derived.  The CE2 model has been described elsewhere\cite{srinivasan:2012,farrell:2003,marston:2008} and will be only briefly reviewed here.  After making the QL approximation discussed in the Introduction, one can form equations for the two-point covariance $W \defineas \langle \zeta \zeta \rangle$ and the zonal flow $U = \langle\unit{x} \cdot \v{v}\rangle$, where angle brackets denote a zonal average.  Recently, an alternative derivation and formulation has been presented that makes use of the Wigner--Moyal formalism.\cite{ruiz:2016}  In that derivation, the CE2 equations of motion are written in phase space as
	\begin{subequations}
	\begin{align}
		\partial_t W &= \{\{ H, W \} \} + [[\G, W]] + F - 2 \m W, \label{CE2_WM_W}\\
		\partial_t U &= -\m U + \partial_y \int \frac{d\vk}{(2\pi)^2} \frac{1}{\kbsq} \star k_x k_y W \star \frac{1}{\kbsq},
	\end{align}
	\end{subequations}
where $W$ now plays the role of the Wigner function, $\kbsq = k^2 + 1$, $\mathcal{H} = -\b k_x / \kbsq + k_x U + [[ U'', k_x / \kbsq]]/2$ is a wavepacket Hamiltonian, $\Gamma = \{\{ U'', k_x / \kbsq\}\}/2$ is an interaction term, and $F$ is the covariance of the white-noise forcing $f$.  For details on the formalism, including on the Moyal sine bracket $\{\{ \cdot, \cdot \} \}$, the cosine bracket $[[ \cdot, \cdot ]]$, and the Moyal product $\star$, the reader is referred to Ref.~\onlinecite{ruiz:2016}.  Our Fourier transform convention is $f(k) = \int dx\, e^{-ikx} f(x)$.  For simplicity, we omit viscosity from the equations here.  It is not difficult to include, and hyperviscosity is used in the simulations discussed later.

From CE2, one can derive the CE2-GO model straightforwardly using a geometrical-optics expansion based on an assumption of separation of spatial scales between the zonal flow and fluctuations.  A separation of timescales is not assumed.  An elementary derivation was presented in Ref.~\onlinecite{parker:2016}.  The reduction can also be carried out within the Wigner--Moyal formalism.\cite{ruiz:2016}  One benefit of this latter derivation is the manifest Hamiltonian behavior; for example, a Moyal bracket reduces to a Poisson bracket in the GO limit.

The CE2-GO equations can be written explicitly as\cite{parker:2016}
	\begin{subequations}
	\label{CE2GO_eqns}
	\begin{align}
		\partial_t W  & -k_x U' \pd{W}{k_y} - k_x U''' \pd{}{k_y} \left( \frac{W}{\kbsq}\right) \notag \\
		  &+ 2(\b - U'') \frac{k_x k_y}{\kbar^4} \pd{W}{y}= F - 2\m W \label{CE2GO_W}, \\
		\partial_t U(y,t) &= -\m U + \partial_y \int \frac{d\vk}{(2\pi)^2} \frac{k_x k_y}{\kbar^4} W. \label{CE2GO_U}
	\end{align}
	\end{subequations}
The GO approximation greatly simplifies the mathematical structure of the CE2 equations.  In phase-space coordinates, the CE2 equation \eref{CE2_WM_W} involves complicated convolutions in the operations of the brackets.  In double-physical-space coordinates rather than phase space, the terms have a simpler form involving only multiplication, such as $[U(y + s_y/2) - U(y - s_y/2)] W(s_x, s_y, y)$, but now there is \emph{nonlocality} appearing in the argument of the zonal flow $U$ in a form typical of Wigner equations.  To obtain the CE2-GO form, Taylor expand in small $s_y$ and Fourier transform from $\v{s} \to \vk$, and this term becomes $U'(y) \partial W(\vk, y) / \partial k_y$, which is a \emph{local} interaction in the phase space.

\subsection{Energy and enstrophy conservation}
In CE2-GO, the energy density can be decomposed into contributions from the eddies and the zonal flow, given by
	\begin{subequations}
	\begin{align}
		E_{e} &= \frac{1}{2L_y} \int_0^{L_y} dy\, \int \frac{d\vk}{(2\pi)^2} \frac{W(\vk,y)}{\kbsq}, \\
		E_{zf} &= \frac{1}{2L_y} \int_0^{L_y} dy\, U(y)^2,
	\end{align}
	\end{subequations}
where $L_y$ is a domain size with periodic boundary conditions assumed.  Similarly, the enstrophy density  is given by
	\begin{subequations}
	\begin{align}
		Z_{e} &= \frac{1}{2L_y} \int_0^{L_y} dy\, \int \frac{d\vk}{(2\pi)^2}W(\vk,y), \\
		Z_{zf} &= \frac{1}{2L_y} \int_0^{L_y} dy\, U'(y)^2.
	\end{align}
	\end{subequations}
In the absence of forcing and dissipation, it is straightforward to show that the total energy $E = E_e + E_{zf}$ and total enstrophy $Z = Z_e + Z_{zf}$ are conserved.  In the limit of extremely-long-wavelength zonal flows in which $U''$ and $U'''$ are neglected, CE2-GO reduces to the traditional WKE, in which the total enstrophy is not conserved.\cite{ruiz:2016}  The necessity of the higher derivatives $U''$ and $U'''$ indicates the effect of zonal flow on fluctuations cannot be solely described by a local shear.

\subsection{Zonostrophic Instability in CE2-GO}
In the presence of incoherent fluctuations, zonal flows can form spontaneously in a symmetry-breaking instability known as zonostrophic instability.\cite{srinivasan:2012,parker:2016}  Here, we recall the basic steps to find the dispersion relation for the instability within the CE2-GO system.

The instability is analyzed by considering an equilibrium consisting of a state of statistically homogeneous fluctuations or turbulence.  In a statistically homogeneous situation, statistical quantities such as the covariance $W = W_H(\vk)$ do not depend on the coordinate $y$, where the $H$ subscript denotes homogeneous.  In this homogeneous state, there is no zonal flow, $U=0$.  Balancing forcing with dissipation, one can find the homogeneous equilibrium, $W_H = F / 2\m$.

Then, a small, symmetry-breaking perturbation with zonal flow is considered:
	\begin{subequations}
	\label{perturbation_forms}
	\begin{align}
		W(\vk, y, t) &= W_H + W_1(\vk) e^{iqy} e^{\l t}, \\
		U(y, t) &= U_1 e^{iqy} e^{\l t}.
	\end{align}
	\end{subequations}
Using Eq.~\eref{perturbation_forms} and linearizing the CE2-GO equations about the homogeneous state, one obtains
	\begin{subequations}
	\begin{align}
		\l W_1 &- ik_x q U_1 \pd{W_H}{k_y} + ik_x q^3 U_1 \pd{}{k_y} \left( \frac{W_H}{\kbsq} \right) \notag \\
			&+ \frac{2i\b q k_x k_y }{\kbar^4} W_1 = -2\m W_1, \label{linearizedeqn_W1} \\
		\l U_1 &= -\m U_1 + iq \int \frac{d\vk}{(2\pi)^2} \frac{k_x k_y}{\kbar^4} W_1. \label{linearizedeqn_U1}
	\end{align}
	\end{subequations}
Equation \eref{linearizedeqn_W1} can be solved for $W_1$ in terms of $U_1$:
	\begin{equation}
		W_1 = iqk_x U_1 \frac{\pd{}{k_y} \left[ \left( 1 - \frac{q^2}{\kbsq} \right) W_H \right]}{\l + 2\m + 2i\b q k_x k_y / \kbar^4}.
	\end{equation}
This relation can be substituted back in to Eq.~\eref{linearizedeqn_U1} to obtain a nonlinear equation for the eigenvalue $\l$,
	\begin{align}
		\l + \m = -q^2 \int \frac{d\vk}{(2\pi)^2} & \frac{k_x^2 k_y}{(\l + 2\m)\kbar^4 + 2i \b q k_x k_y} \notag \\
			& \times \pd{}{k_y} \left[ \left( 1 - \frac{q^2}{\kbar^2} \right) W_H \right], \label{nonlinear_equation_lambda}
	\end{align}
which is the dispersion relation for zonostrophic instability within CE2-GO.  Given a functional form for $W_H$, Eq.~\eref{nonlinear_equation_lambda} can be solved numerically for $\l$.  Unstable eigenvalues are typically real, although in certain cases exceptions can exist.\cite{ruiz:2016}  Real eigenvalues imply the zonal flow perturbation grows in place rather than propagating as a wave.

\section{Conservative numerical simulation of CE2-GO}
\label{sec:numerical_simulation}
\subsection{Pseudospectral simulation}
We present a simulation of the CE2-GO model that is spectral in $k_x$ and pseudospectral in $k_y$ and $y$.  Periodic boundary conditions are assumed.  

In the pseudospectral procedure, we denote the Fourier variable conjugate to $y$ as $q$ and the Fourier variable conjugate to $k_y$ as $s_y$.  Derivatives in $y$ are computed in Fourier space by multiplying by $iq$.  Moreover, the derivatives in $k_y$ are also computed pseudospectrally: quantities are Fourier transformed to their conjugate physical space, multiplied by $-is_y$, and then transformed back to $k_y$ space.  The quadratic products in Eq.~\eref{CE2GO_W} involving the zonal flow multiplying the covariance is carried out by Fourier transforming both quantities from Fourier space $q$ to physical space $y$, performing the multiplication, and transforming back to Fourier space.  

In Ref.~\onlinecite{ruiz:2016}, a single simulation of CE2-GO was presented under the name ``WKE.''  That simulation was based on a discontinuous-Galerkin finite-element method and differs from the conservative pseudospectral scheme presented here.

\subsection{Energy and enstrophy conservation in numerical simulation}
In ordinary pseudospectral simulation of the (non-statistically averaged) Hasegawa--Mima equation, exact conservation of the quadratic invariants, energy and enstrophy, can be achieved by 2/3 dealiasing.\cite{orszag:1971:dealias,boyd:2001}  The dealiasing procedure for quadratic nonlinearities  can be adapted straightforwardly to the nonlinear pseudospectral products in the CE2-GO equation \eref{CE2GO_W}.  No dealiasing is required in the Reynolds-stress term on the right-hand-side of Eq.~\eref{CE2GO_U} because it is linear in $W$.  In the Hasegawa--Mima equation, the corresponding term is nonlinear in $\z$.

With dealiasing, one can achieve exact energy and enstrophy conservation in the pseudospectral formulation, but some care is required.  Derivation of this conservation from Eq.~\eref{CE2GO_eqns} shows that cancellation of terms between $\dot{W}$ and $\dot{U}$ relies on 1) an integration by parts in $k_y$ and 2) that $\partial_{k_y} (1/\kbsq) = -2 k_y / \kbar^4$.    The problem with the form $(k_y / \kbar^4)$ is that in the \emph{discrete} Fourier transform, where derivatives are computed pseudospectrally as described above, it is \emph{not true} that $\partial_{k_y} (1/\kbsq) = -2 k_y / \kbar^4$.  Integration by parts, however, still holds in the discrete Fourier transform, meaning for any $f$, $\sum_{k_y} f_{k_y} \partial W_{k_y} / \partial k_y = - \sum_{k_y} (\partial f_{k_y} / \partial k_y) W_{k_y}$.  We can thus rewrite the CE2-GO equations in conservative form:
	\begin{subequations}
	\label{CE2GO_conservative}
	\begin{align}
		\partial_t W &+ \pd{\w}{k_y} \pd{W}{y} - \pd{}{k_y} \left[ \pd{\w}{y} W \right] = F - 2\m W,  \label{CE2GO_W_conservative} \\
		\partial_t U &+ \m U = -\partial_y \int \frac{d\vk}{(2\pi)^2} \frac{k_x}{2} \pd{}{k_y} \left( \frac{1}{\kbsq} \right) W(\vk, y), \label{CE2GO_U_conservative}
	\end{align}
	\end{subequations}
where $\w(\vk, y) = -k_y[\b - U''(y)]/\kbsq + k_x U(y)$ is the quasilinear wave frequency and is identical to the GO limit of the Hamiltonian $\mathcal{H}$.  In Eq.~\eref{CE2GO_U_conservative}, the integral is discretized $(2\pi)^{-2} \int d\vk\, f(\vk) \to \sum_\vk f_\vk$.

Now, the form of Eq.~\eref{CE2GO_conservative} numerically maintains the nonlinear invariants because the integration-by-parts property of the pseudospectral derivative is sufficient to guarantee conservation.  Dealiasing using the $2/3$ rule for the nonlinear terms in Eq.~\eref{CE2GO_W_conservative} is required.  In this manner, $dE/dt=0$ and $dZ/dt=0$ can be achieved.  However, as is typical, some error in conservation is introduced by temporal discretization.

\begin{figure}
		\centering
		\includegraphics{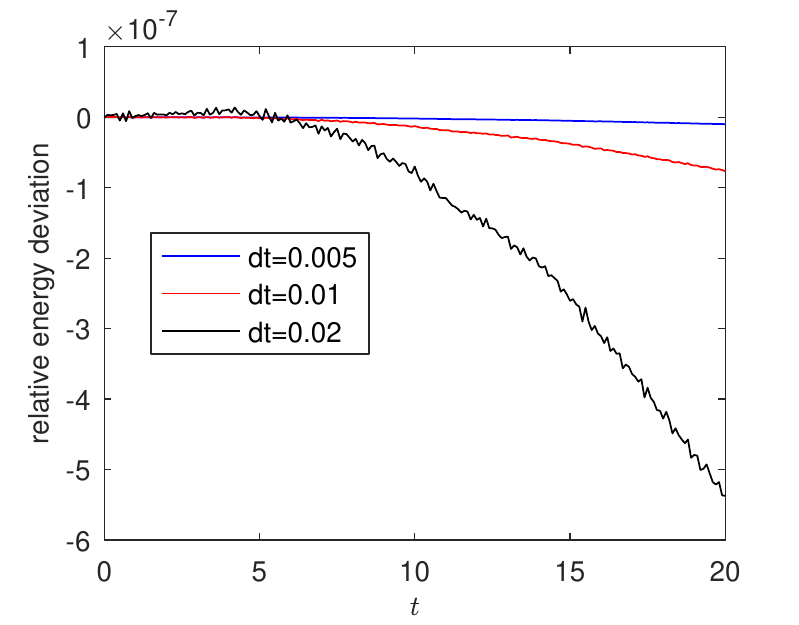}
		\caption{Energy conservation in the pseudospectral CE2-GO simulations for several timesteps $dt$ when forcing and dissipation are turned off.  The plot shows $[E(t) - E(0)]/E(0)$.  A third-order RK3CN timestepper is used.  The corresponding plot for enstrophy looks much the same.  At each timestep, the instantaneous changes in total energy and enstrophy, $\dot{E}$ and $\dot{Z}$, are approximately zero to machine precision.}
		\label{fig:ce2go_energy_conservation}
	\end{figure}

The energy and enstrophy conservation can be illustrated with nonlinear simulations in which forcing and dissipation are set to zero.  These simulations use a semi-implicit RK3CN timestepper.\cite{lundbladh:1999}  At each time, the instantaneous relative change in energy, $\dot{E} / \dot{E}_{zf}$ is smaller than $10^{-14}$, and similarly for enstrophy.  Figure \ref{fig:ce2go_energy_conservation} illustrates the energy conservation as a function of time.  The plot shows the relative change in total energy $[E(t) - E(0)] / E(0)$ as a function of time, for three different timestep sizes.  The corresponding plot for enstrophy looks much the same.  These results demonstrate that non-conservation error is introduced only by finite machine precision and discrete timestepping, and the error can be kept small.

In the following section, we compare the CE2-GO simulations with direct (non-statistical) simulations of the QL dynamics.  The QL simulations are standard, using periodic boundary conditions and pseudospectral methods with dealiasing.

\section{Results}
\label{sec:results}
Using the conservative form of CE2-GO, we perform nonlinear simulations and compare to QL dynamics in order to assess the fidelity of the GO approximation.    We use parameters that have been previously studied in investigations of CE2 and QL dynamics.\cite{srinivasan:2012,parker:2013,parker:2014,parker:2016}  The study of this regime in the context of the GO approximation is novel.

The parameters are as follows: the external forcing is given as a ring in wavenumber space, with covariance $F(\vk) = 4\pi \ve k_f \de(k - k_f)$ (the delta function is discretized as a thin, finite-width ring).  We take $\b = 1$, $\ve=1$, $\nu = 3 \times 10^{-4}$ with eighth-order hyperviscosity, and vary $\m$.  We use $(N_{k_x}, N_{k_y}, N_y) = (16, 48, 400)$ points, with a spectral resolution $\D k_x = 0.15$, $\D k_y = 0.15$.  Denoting the Fourier variable conjugate to $y$ by $q$, we use spectral resolution $\D q=0.04$.  Convergence has been checked by halving each of $\D k_x, \D k_y, \D q$ in select instances.

Our QL simulations are also pseudospectral.  We use $(N_x, N_y) = (256, 256)$ points, with a spectral resolution $\D k_x = 0.01$ and $\D k_y = 0.04$.  The reason $\D k_x$ is smaller is to increase the domain size $L_x = 2\pi / \D k_x$, which serves to bring the system closer to ergodicity where a zonal average over $x$ is equivalent to an ensemble average.  Because the QL approximation results in fluctuations not scattering to higher $k_x$, the maximum resolved wavenumber $N_x \D k_x/2$ need not be very large.  The domain size $L_y$ is chosen to allow for many wavelengths of zonal flow to fit into the system, as too small a domain could affect the outcome.  

\begin{figure}
		\centering
		\includegraphics[width=3.2in]{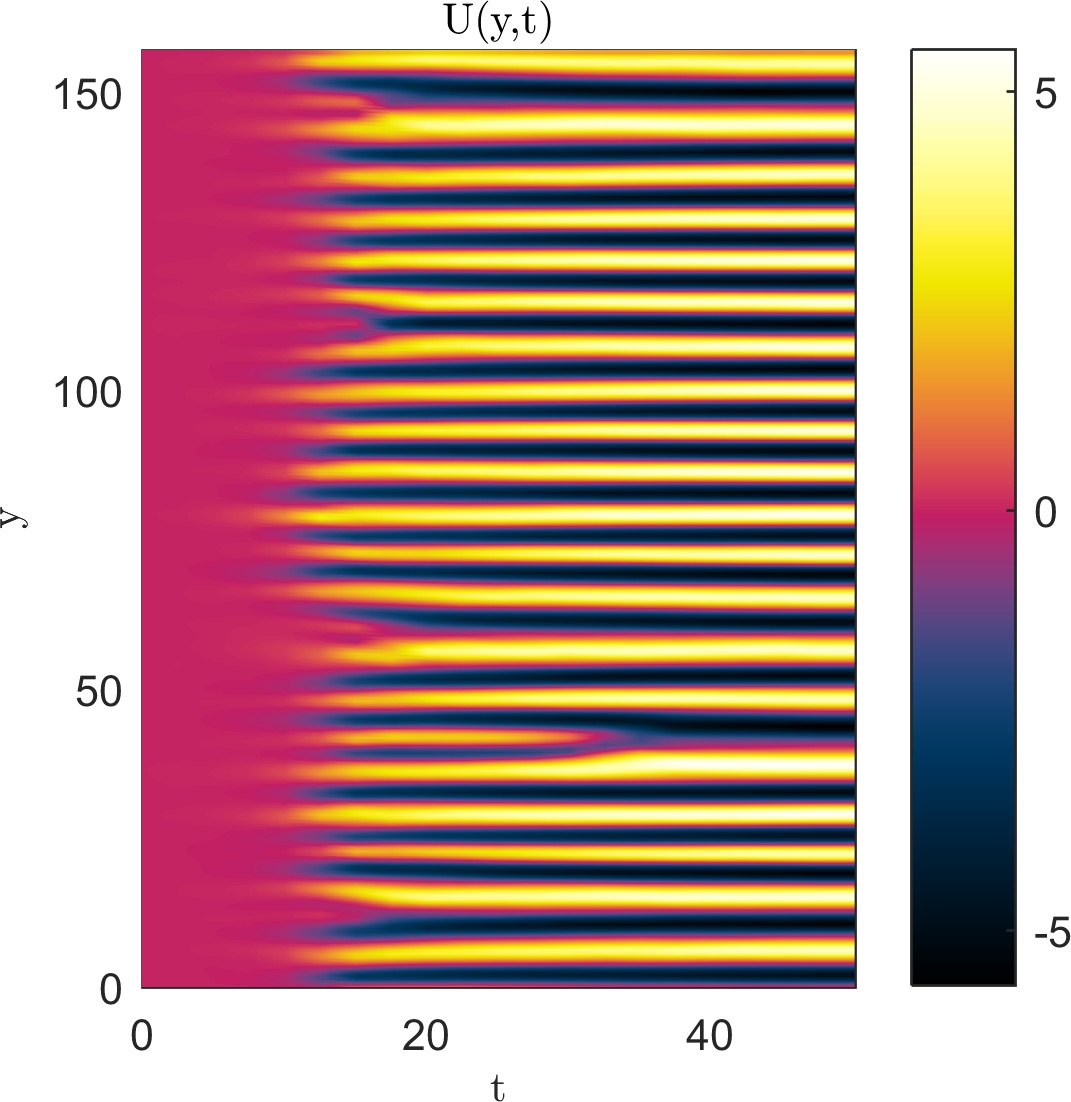}
		\caption{Zonal flow $U(y,t)$ as a function of space and time in the QL simulation.}
		\label{fig:ql_zfspacetime}
	\end{figure}

\begin{figure}
		\centering
		\includegraphics[width=3.2in]{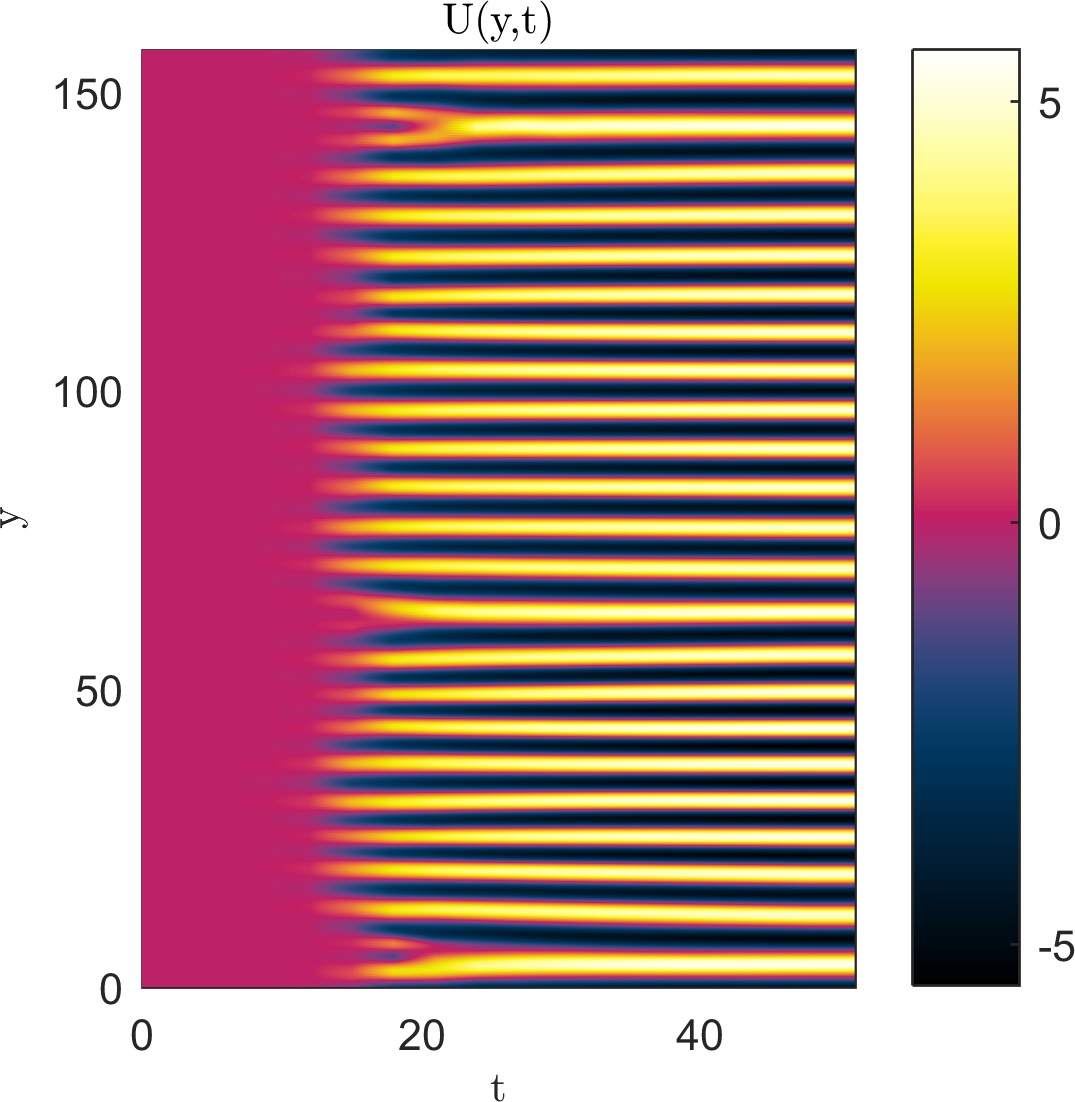}
		\caption{Zonal flow $U(y,t)$ as a function of space and time in the CE2-GO simulation.}
		\label{fig:ce2go_zfspacetime}
	\end{figure}

\begin{figure}
		\centering
		\includegraphics[width=3.2in]{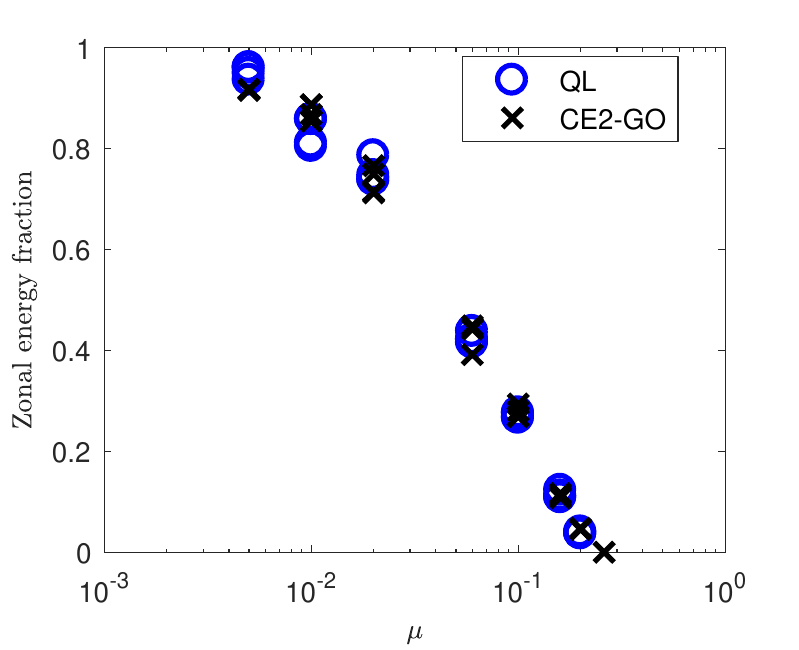}
		\caption{Fraction of the energy in the zonal flow in the steady state as a function of the dissipation coefficient $\m$ in QL and CE2-GO simulations.  Points represent individual simulations starting from four different initial conditions at each parameter value for both CE2-GO and QL.}
		\label{fig:zf_energy_fraction}
	\end{figure}

In Figures \ref{fig:ql_zfspacetime} and \ref{fig:ce2go_zfspacetime}, we show the zonal flow behavior in space and time at $\m = 0.02$ at early times during the transient behavior of the QL and CE2-GO simulations.  Mergings of jets can be seen in both plots.  In Figure \ref{fig:zf_energy_fraction}, we show the fraction of energy in the zonal flow $E_{zf} / E$ after the steady state is reached as a function of $\m$.  The values for four simulations at each parameter value are shown to capture the variation due to different initial conditions and (for QL) different realizations of the random forcing.  At large $\m$, between 0.2 and 0.26, there is a critical onset value, above which the zonal flow energy in CE2-GO is zero and there is no zonal flow.  For $\m$ below the onset value, a zonal flow forms.  The steady-state energy fraction of the zonal flow agrees well quantitatively between QL and CE2-GO, increasing towards 1 as $\m$ decreases.  

\begin{figure}
		\centering
		\includegraphics{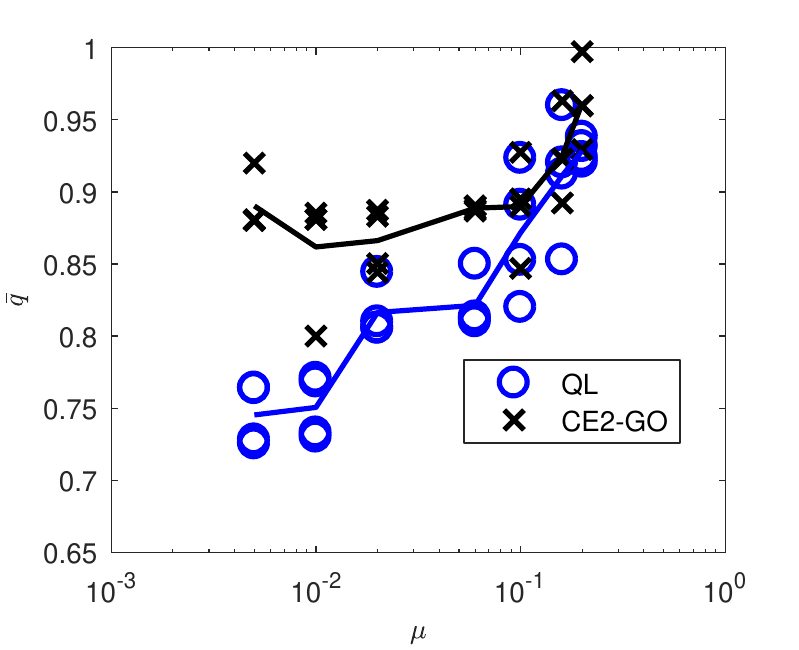}
		\caption{Energy-weighted mean wavenumber $\ol{q}$ of the zonal flow, defined in Eq.~\eref{energy_weighted_wavenumber}.  Points represent individual simulations starting from four different initial conditions at each parameter value for both CE2-GO and QL.  The lines have been added as the mean value of $\ol{q}$ over the simulations at each $\m$ value.}
		\label{fig:q_vs_mu}
	\end{figure}

In Figure \ref{fig:q_vs_mu}, we plot the energy-weighted mean wavenumber $\ol{q}$ of the zonal flow, defined by
	\begin{equation}
		\ol{q} \defineas \frac{ \sum_{q} |U_{q}|^2 q}{\sum_{q} |U_{q}|^2},
		\label{energy_weighted_wavenumber}
	\end{equation}
where the sum is over positive $q$.  The values for four simulations at each parameter value are shown, with a line added to show the mean value over the four simulations.  If one counts the number of wavelengths or bands $N$ in the system, then $N \Delta q$ in CE2-GO (or N$\D k_y$ in QL) is very close to $\ol{q}$.  In the QL simulations, the mean wavenumber $\ol{q}$ shows a clear trend of decreasing as $\m$ decreases from the onset value, reaching $\ol{q} \approx 0.75$ at $\m=0.005$.  In contrast, in the CE2-GO simulations, there is a small drop in $\ol{q}$ around $\m=0.1$, close to the onset, but as $\m$ gets smaller, $\ol{q}$ does not change much, with $\ol{q} \approx 0.88$ at $\m=0.005$.

\section{Discussion}
\label{sec:discussion}
The onset of zonal flow in a supercritical bifurcation at the marginal point of zonostrophic instability occurs here for $\m$ between 0.2 and 0.26.  As this critical value for the transition to zonal flow agrees for CE2-GO and QL, we conclude that this behavior in CE2-GO is inherited faithfully from CE2.  This is unsurprising, as we have seen how the dispersion relation for zonostrophic instability in CE2-GO agrees well with that of CE2.\cite{parker:2016}.  Based on Figure \ref{fig:zf_energy_fraction}, in which there is good quantitative agreement in the steady-state zonal-energy fraction, we also conclude that the mechanisms for energy transfer between fluctuations and zonal flow persist in the GO reduction from CE2 to CE2-GO.

In both fully nonlinear and QL simulations, it has long been observed in the barotropic vorticity equation that the length scale of zonal flow tends to get larger as the energy in the flow increases.  The barotropic vorticity equation, which has the same form as the Modified Hasegawa--Mima equation except Eq.~\eref{HMPoissonEqn} is replaced by $\z = \nabla^2 \psi$, has been studied extensively.\cite{vasavada:2005,vallis:1993}

The Rhines scale $L_R = \sqrt{U / \b}$ has been found to often provide good approximate agreement with the observed length scale of the zonal flow, where $U$ is the root-mean-square (rms) flow velocity.\cite{rhines:1975}  A physical description of the Rhines scale arises from calculating the length scale at which the inertial frequency associated with the inverse cascade becomes comparable with the Rossby frequency.  For constant energy input, as the dissipation $\m$ decreases, the rms velocity $U$ will increase, which implies an increase in the size of the Rhines scale.  Therefore, an increase in length scale of zonal flow (or a decrease in wavenumber) is the expected behavior as $\m$ decreases.  That is what is observed in the QL simulations, as seen in Figure \ref{fig:q_vs_mu}.  In contrast, in the CE2-GO simulations, $\ol{q}$ saturates as $\m$ decreases, suggesting a Rhines-like scaling is not obeyed.  Whatever mechanism is responsible for the increase in characteristic scale of the zonal flow in the QL simulations, these simulations suggest that the relevant physics may be lost in the GO approximation.

A connection between the stability boundary for steady states of zonal flows and the final equilibrated length scale was discussed in Refs. \onlinecite{parker:2013} and \onlinecite{parker:2014} in the context of CE2.  This boundary describes the stability of \emph{finite amplitude} zonal flows.  This is distinct from the zonostrophic instability, which describes the growth of infinitesimal zonal flows.  In those works, it was argued that for some parameters far beyond the onset value where zonal flows first form, the zonostrophic instability causes growth of zonal flows which have higher wavenumber than can ultimately be stably equilibrated.  To reach a steady state, the zonal flows must then change their wavenumber so as to cross into a stable region.  The mechanism by which this occurs is that a high-wavenumber zonal flow undergoes an instability, the nonlinear consequence of which is the merging of jets, as seen in Figures \ref{fig:ql_zfspacetime} and \ref{fig:ce2go_zfspacetime}.  The crossing of these stability boundaries can be thought of as manifestations of a type of tertiary instability.\cite{rogers:2000,onge:2017}  Close to the onset value of $\m$, the broken symmetry of translational invariance in the $y$ direction guarantees that the Eckhaus instability will be active.\cite{parker:2014}  Far from the onset value, however, is where other instabilities that set the boundary may be lost in the GO reduction.

Our result is similar to the conclusion of recent work, which proposed that the GO approximation eliminates the tertiary instability for short-wavelength ZF.\cite{zhu:2018}  The loss of the physical processes by which zonal flows equilibrate to larger scales would be a major impediment to the practical use of CE2-GO.  On the other hand, further study of what is lost in the GO approximation may provide greater insight into the tertiary instability.

\section{Conclusion}
\label{sec:conclusion}
A complete theory of zonal flow has yet to be achieved.  One promising approach follows a statistical methodology, based on a quasilinear approximation.  The purpose of this work is to investigate the statistical CE2-GO model numerically.  CE2-GO, a geometrical-optics reduction of the more complete model CE2, has previously been shown to agree well with exact quasilinear dynamics in the dispersion relation for growth of small zonal flow.\cite{parker:2016} In this work, we go further and perform nonlinear simulations of CE2-GO to compare with quasilinear dynamics and assess the fidelity of the GO approximation.

In summary, we have carried out nonlinear simulations of CE2-GO and investigated the behavior of saturated zonal flows.  We have been careful to formulate the numerics in a way to preserve nonlinear conservation of energy and enstrophy.  Our simulations show that some qualitative trends in QL are captured by the CE2-GO model, such as the partitioning of energy between zonal flow and fluctuations.  On the other hand, because the CE2-GO and QL simulations exhibit differing qualitative trends of the steady-state length scale of the zonal flow, the GO approximation appears to lose important physics.  Further study of what is lost in the GO approximation may provide a greater understanding of the processes such as tertiary instability that set the zonal flow scale.

\begin{acknowledgments}
Useful discussions with Ilya Dodin, Daniel Ruiz, and Eric Shi are acknowledged.  This work was performed under the auspices of the U.S.\ Department of Energy by Lawrence Livermore National Laboratory under Contract No.\ DE-AC52-07NA27344.
\end{acknowledgments}

\bibliographystyle{apsrev4-1}
\bibliography{ce2go_paper_arxiv} % if running BiBTeX	
	
\end{document}